\documentclass[aps,pre]{revtex4}
\usepackage{amsmath,amssymb,amsthm,mathrsfs,amsfonts,dsfont} 
\usepackage{graphicx}
\usepackage{subfigure}
\usepackage{amsbsy}
\usepackage{bm}

\def\bs{\mathbf{s}}
\def\bx{\boldsymbol{\xi}}
\def\X{\mathbf{X}}

\def\d{\mathrm{d}}
\def\J{\mathbf{J}}
\def\I{\mathbf{I}}
\def\tr{\operatorname{Tr}}
\def\det{\operatorname{det}}
\def\im{\operatorname{Im}}
\def\xih{\hat{\xi}}
\def\M{\mathcal{M}}
\def\Q{\mathcal{Q}}

\def\q{\tilde{q}}

\def\t{\operatorname{T}}

\def\ld{\left\langle}
\def\rd{\right\rangle}

\def\l{\left}
\def\r{\right}

\def\bxih{\boldsymbol{\hat{\xi}}}

\def\bX{\mathbf{X}}
\def\bx{\boldsymbol{\xi}}

\def\lam{\lambda}
\def\f{\frac}

\def\i{\mathrm{i}}
\def\e{\epsilon}

\def\I{\mathbf{I}}

\def\bJ{\mathbf{J}}
\def\bm{\mathbf{m}}

\def\Tr{\operatorname{Tr}}
\def\p{\rho}

\def\T{\mathtt{T} }
\def\a{\alpha}

\def\bxi{\boldsymbol{\xi}}
\def\xih{\hat{\xi}}

\def\bLam{\mathbf{\Lambda}}
\def\bO{\mathbf{O}}

\def\rank{\operatorname{rank}}

\begin{document}

\title{Eigenvalue spectrum of neural networks with arbitrary Hebbian length}
\author{Jianwen Zhou}
\thanks{J.Z., Z.J. and T.H. contributed equally to this work.}
\affiliation{PMI Lab, School of Physics,
Sun Yat-sen University, Guangzhou 510275, People's Republic of China}
\author{Zijian Jiang}
\thanks{J.Z., Z.J. and T.H. contributed equally to this work.}
\affiliation{PMI Lab, School of Physics,
Sun Yat-sen University, Guangzhou 510275, People's Republic of China}
\author{Tianqi Hou}
\thanks{J.Z., Z.J. and T.H. contributed equally to this work.}
\affiliation{Department of Physics, the Hong Kong University of Science and Technology, Clear Water Bay, Hong Kong}
\affiliation{Theory Lab, Central Research Institute, 2012 Labs, Huawei Technologies Co., Ltd.}
\author{Ziming Chen}
\affiliation{PMI Lab, School of Physics,
Sun Yat-sen University, Guangzhou 510275, People's Republic of China}
\author{K. Y. Michael Wong}
\affiliation{Department of Physics, the Hong Kong University of Science and Technology, Clear Water Bay, Hong Kong}
\author{Haiping Huang}
\email{huanghp7@mail.sysu.edu.cn}
\affiliation{PMI Lab, School of Physics,
Sun Yat-sen University, Guangzhou 510275, People's Republic of China}
\date{\today}
\begin{abstract}
Associative memory is a fundamental function in the brain. Here, we generalize the standard associative memory model to include long-range Hebbian interactions at 
the learning stage, corresponding to a large synaptic integration window. In our model, the Hebbian length can be arbitrarily large. The spectral density of the coupling
matrix is derived using the replica method, which is also
shown to be consistent with the results obtained by applying the free probability method. The maximal eigenvalue is then obtained by an iterative equation, related to 
the paramagnetic to spin glass transition in the model. Altogether, this work establishes the connection between the associative memory with arbitrary Hebbian length and 
the asymptotic eigen-spectrum of the neural-coupling matrix.
\end{abstract}

 \maketitle

\section{Introduction}
The ability for a neural circuit to associate a cue to a target memory is fundamental to the brain across species.
The associative memory has been modeled as a physics system~\cite{Amit-1989}, and observed in neural circuits (e.g., Hippocampal networks~\cite{Fusi-2020,Guzman-2016}).
Recently, we propose an associative memory model with arbitrary Hebbian length~\cite{Huang-2021}. More precisely, when constructing the synaptic coupling between a pair of neurons
in a neural network, Hebbian learning occurs at non-neighboring patterns (e.g., separated by as far as $d$ patterns), provided that these patterns form a cyclic sequence for the model to 
learn. As already revealed, the model taking into account Hebbian learning of only neighboring patterns (i.e., $d=1$) is able to
convert the temporal correlations in the ordered stored sequence into the spatial correlations of attractors of neural activity~\cite{Amit-1993,Tsodyks-1994}.
Interestingly, by taking into account one additional pattern separation (i.e., $d=2$), the correlated attractor phase can be significantly enhanced in terms of correlation spans~\cite{Huang-2021}.
A longer Hebbian length (i.e., $d$) corresponds to a wider learning window, which has been observed in different neural circuits, e.g., spiking-time-dependent synaptic plasticity~\cite{Bittner-2017,Hebbian-2018,Hebbian-2020}.
Therefore, the model of arbitrary Hebbian length is important to address how the synaptic integration time window (at the
microscopic level) affects the global attractor properties (e.g., emergence of correlated attractor phase with different correlation spans).

The phase diagram of the model was recently analytically studied~\cite{Huang-2021} by using replica method in disordered spin glass theory~\cite{Mezard-1987}.
The equilibrium properties and even non-equilibrium dynamics towards the stationary state guaranteed by the symmetric coupling setting are related to the eigen-spectrum of the coupling matrix.
It is thus essential to explore the asymptotic eigenvalue density of the model. In particular, the maximal eigenvalue is related to the linear stability of the paramagnetic phase, determining the transition to 
the spin glass phase~\cite{JPA-1997,Kabashima-2010,Mezard-2017}. In this paper, we apply the replica method to derive the eigen-spectrum, whose analytic form is also confirmed by 
the free probability theory~\cite{FP-2017}. We also explore the effects of model parameters, such as memory load, Hebbian strength, and Hebbian length, on the eigen-spectrum and paramagnetic-to-spin-glass transition.
The theoretical predictions are in an excellent agreement with the numerical simulations. Comparison to the standard Hopfield model provides insights about the significance of our model, and implications to the temporal association memory 
in neural circuits.

\section{Model}
We study a recurrent neural network composed of $N$ interacting neurons.
The synaptic coupling between any two neurons (say $i$ and $j$) is given by
\begin{equation}
	J_{ij} = \frac{1}{N} \sum_{\mu=1}^P\left[c \xi_{i}^{\mu} \xi_{j}^{\mu}+\gamma \sum_{r=1}^{d}\left(\xi_{i}^{\mu} \xi_{j}^{\mu+r}+\xi_{i}^{\mu+r} \xi_{j}^{\mu}\right)\right]\ ,
\end{equation}
which combines a non-delayed (concurrent) Hebbian term and an arbitrarily-delayed (non-concurrent) Hebbian term at the reciprocal connection. The strengths of these two terms are specified by $c$ and $\gamma$, respectively.
The delay means that the patterns used to construct the Hebbian coupling can be shifted with a time interval, provided that all patterns stored in the network form a cyclic sequence.
The time interval (delay) is called the Hebbian length in our model, denoted as $d$. The Hamiltonian of the model reads as follows,
\begin{equation}\label{eq2}
 \mathcal{H}(\bs)=-\frac{1}{2}\sum_{i\neq j}J_{ij}s_is_j.
\end{equation}

The simplest case of $d=0$ corresponds to the standard Hopfield model~\cite{Amari-1977,Hopfield-1982}, while another simple case of $d=1$ has been used to model the correlated spatial
attractors observed in the temporal cortex of the Monkey~\cite{Miya-1988a,Miya-1988b,Amit-1993,Tsodyks-1994,Fukai-2019}. Although the memory attractors in the recall phase are independent fixed points, 
by tuning the other model parameters (e.g., $\gamma$ or $d$), there appears a correlated attractor phase where the network stimulated by each of the stored patterns evolves to an attractor, which correlates with a small number 
of patterns concentrating around the stimulating pattern. In other words, the correlation between two attractors triggered by two stimulating patterns depends only on
the separation of the corresponding stimulus in the cyclic sequence. Then, the correlation span can be precisely controlled by modifying the microscopic details of synaptic couplings.

In our model setting, the cyclic sequence has $P$ patterns. Their entries $\xi_{i}^\mu$ where $\mu$ is the pattern index and $i$ is the site index
follows an independent 
binary distribution, i.e., $p(\xi_{i}^\mu=\pm 1)=\f{1}{2} \delta(\xi_i^\mu +1)+\f{1}{2} \delta(\xi_i^\mu -1)$.
We are interested in the regime of large values of $P$ and $N$, thereby defining the memory load
\begin{equation}
	\alpha = \frac{P}{N}\ ,
\end{equation}
and the coupling matrix $\bJ$ can be written as
\begin{equation}
	\bJ =  \frac{1}{N} \bxi^\T \bX \bxi ,
\end{equation}
where $\bX$ is a $P\times P$ circulant matrix, a special form of Toeplitz matrix with elements
\begin{equation}
	X_{\mu\nu}=c\delta_{\mu\nu}+\gamma\sum_{r=1}^d(\delta_{\mu,(\nu-r)\,{\rm mod}\,P}+\delta_{\mu,(\nu+r)\,{\rm mod}\,P}).
\end{equation}
For example, for $d=1$, and $P=5$, $\mathbf{X}$ reads
 \begin{equation}
 \left(
 \begin{matrix}
c&\gamma&0&0&\gamma\\
\gamma&c&\gamma&0&0\\
0&\gamma&c&\gamma&0\\
0&0&\gamma&c&\gamma\\
\gamma&0&0&\gamma&c\\
 \end{matrix}
 \right).
 \label{eq:example}
 \end{equation}
Note that the number of $\gamma$ in each row is conserved.
The $m$-th eigenvalue of $\bX$ is given by~\cite{Gray-2005}
\begin{equation}\label{eigX}
	\Lambda_{m} =
c  +  2 \gamma \sum_{r=1}^d  \cos\l({2\pi \frac{mr }{P}} \r) \ ,
\end{equation}
for $m= 0, 1,\cdots, P-1$. 

The circulant matrix $\bX$ can be diagonalized by an orthogonal matrix $\bO$ as $\bX=\bO^\T \bLam \bO$, 
where $\bLam=\operatorname{diag}(\Lambda_0,\cdots,\Lambda_{P-1})$. 
Because an orthogonal transform preserves the statistics of $\bx$,
we introduce rotated memory patterns, $\bxih_i=[\xih_i^1,\cdots,\xih_i^P]^\T $, as
\begin{equation}
	\bxih_i = \bO \bxi_i  \, .
\end{equation}
Each component is thus given by
\begin{equation}
	\xih_i^\mu = \sum_{\nu=1}^P O^\mu _\nu \xi^{\nu}_i \ .
\end{equation}
In the limit $P\to \infty$, using the central limit theorem, we have
\begin{subequations}
\begin{align}
	&\l\langle \sum_{\nu}^P O^\mu _\nu \xi^{\nu}_i \right\rangle_{\xi_i^\mu} = 0 \ ,\\
	&\l\langle \l( \sum_{\nu=1}^P O^\mu _\nu \xi^{\nu}_i \r)^2\right\rangle_{\xi_i^\mu} = \sum_{\nu=1}^P (O_\nu^\mu)^2 (\xi^{\nu}_i)^2 = \sum_{\nu=1}^P (O_\nu^\mu)^2=1 \ .
\end{align}
	\end{subequations}
We thus conclude that the element of $\bxih$ follows a standard Gaussian distribution.
As a consequence,
the synaptic matrix can be recast into the form as 
\begin{equation}
\label{eq:J}
	\bJ =  \frac{1}{N} (\bO\bxi)^\T \bLam \bO\bxi = \frac{1}{N} \bxih^\T \bLam \bxih\ .
\end{equation}

\section{Eigen-spectrum of the coupling matrix}
The eigenvalue spectrum of $\J$ averaged over the quenched disorder is defined as
\begin{equation}
\rho(\lambda) = \frac{1}{N}\ld\sum_i^N \delta (\lambda-\lambda_i)\rd,
\label{eq:rho_def}
\end{equation}
where $\lambda_i$ is the $i^{\mbox{th}}$ eigenvalue of $\J$, and $\ld\cdotp\rd$ denotes the quenched average over the distribution
of $\boldsymbol{\xih}$. $\rho(\lambda)$ can be understood as the averaged eigenvalue density at $\lambda$. 

To obtain an analytic form of Eq.~(\ref{eq:rho_def}) in practice, we introduce the Stieltjes transform of $\rho(\lambda)$ as follows,
\begin{equation}
\begin{aligned}
G_{\mathbf{J}}(\lambda) &= \frac{1}{N} \ld \mbox{Tr}\frac{1}{\lambda\mathbf{I}-\mathbf{J}}\rd, 
\label{eq:R}
\end{aligned}
\end{equation}
where $\mathbf{I}$ is an identity matrix. $(\lambda\mathbf{I}-\mathbf{J})^{-1}$ denotes the resolvent of $\J$.
The eigenvalue spectrum of $\J$ can be obtained by the inverse Stieltjes transform as
\begin{equation}
\rho(\lambda) = \frac{1}{\pi}\lim_{\epsilon\to0^+}\mbox{Im} G_{\mathbf{J}}(\lambda-\i\epsilon),
\label{eq:rho}
\end{equation}
due to the nice property of the Stieltjes transform, i.e.,  $\lim_{\epsilon\to0^+}G_{\J}(\lambda\pm\i\epsilon)=\mathds{h}(\lambda)\mp\i\pi\rho(\lambda)$,
where $\mathds{h}(\lambda)$ denotes the Hilbert transform of $\rho$~\cite{Clean-2017}.

\subsection{Replica method}
Equation~(\ref{eq:rho}) can be rewritten as follows,
\begin{equation}
\begin{aligned}
\rho(\lambda)
&=\frac{1}{N\pi}\lim _{\epsilon \rightarrow 0^{+}} \operatorname{Im}\frac{\partial}{\partial\lambda}\left<\ln\operatorname{det}(\lambda_\epsilon\mathbf{I}-\J)\right>_{\J}\\
&=\frac{-2}{\pi N} \lim _{\epsilon \rightarrow 0^{+}} \operatorname{Im} \frac{\partial}{\partial \lambda}\langle\ln Z(\lambda_\epsilon)\rangle_\J \ ,
\end{aligned}
\end{equation}
where we have used $G_{\J}(\lambda)=\frac{1}{N}\frac{\partial}{\partial\lambda}\langle\Tr\ln(\lambda\I-\J)\rangle$ and $\operatorname{det}(e^{\mathbf{A}})=e^{\Tr\mathbf{A}}$ ($\mathbf{A}$ is a matrix),
$\lambda_\epsilon=\lambda-\i\epsilon$, and the partition function is defined as~\cite{EJ-1976}
\begin{equation}
\begin{aligned}
Z(\lambda) &\equiv [\det(\lambda\mathbf{I}-\mathbf{J})]^{-\frac{1}{2}}\\
&=\frac{e^{\frac{\i N\pi}{4}}}{(\pi)^{\frac{N}{2}}}\int_{-\infty}^{\infty} \left(\prod_i \d y_i\right) \exp \left[-\frac{\i\lambda}{2}\sum_i (y_i)^2+\frac{\i}{2}\sum_{i,j}y_i J_{ij}y_j\right],
\label{eq:Z}
\end{aligned}
\end{equation}
where $i,j$ goes from $1$ to $N$, marked as the neuron index. 
To derive Eq.~(\ref{eq:Z}), we have applied the multivariate Fresnel integral.

We then apply the replica trick to compute the disorder average over the random matrix ensemble, i.e.,
$\langle \ln Z \rangle = \lim_{n\to 0} \frac{\ln\langle Z^n\rangle}{n}$. 
The eigenvalue spectrum is then explicitly given by
\begin{equation}
\rho(\lambda) = -\frac{2}{N\pi}\lim_{\epsilon\to 0+}\im\frac{\partial}{\partial \lambda} \lim_{n\to 0}\frac{\ln \langle Z^n(\lambda-\i\epsilon) \rangle}{n},
\label{eq:rho_replica}
\end{equation}
where
\begin{equation}
\langle Z^n(\lambda) \rangle 
\propto\int_{-\infty}^{\infty} \d \mathbf{Y} \left\langle\exp \left[-\frac{\i\lambda}{2}\sum_{i,a} (y_i^a)^2+\frac{\i}{2}\sum_{i,j,a}y_i^a J_{ij}y_j^a\right]\right\rangle,
\label{eq:Zn}
\end{equation}
where $\d \mathbf{Y} $ denotes $\left(\prod_{i,a} \d y_i^a\right)$, and $a$ goes from $1$ to $n$, marked as the replica index. 
We have omitted the pre-factor in Eq.~(\ref{eq:Z}) which has no contribution to the eigen-spectrum.
Inserting the explicit form of $\mathbf{J}$ [Eq.~(\ref{eq:J})] into Eq.~(\ref{eq:Zn}), we have:
\begin{equation}
\langle Z^n(\lambda) \rangle 
\propto\int_{-\infty}^{\infty} \d \mathbf{Y} \exp \left[-\frac{\i\lambda}{2}\sum_{i,a} (y_i^a)^2\right]
\times\left\langle\exp\left[\frac{\i}{2N}\sum_{a,\mu}\Lambda_{\mu}\left(\sum_i {\xih}^{\mu}_i y_i^a\right)^2\right]\right\rangle,
\label{eq:Zn1}
\end{equation}
where $\mu$ is the pattern index, going from $1$ to $P$. We now use the Hubbard-Stratonovich transform to linearize the quadratic term
$\left(\sum_i {\xih}^{\mu}_i y_i^a\right)^2$ as
\begin{equation}
\begin{aligned}
&\ld\exp \left[\frac{\mathrm{i}}{2 N} \sum_{a,\mu} \Lambda_{\mu} \left(\sum_{i} \hat{\xi}_{i}^{\mu} y_{i}^{a}\right)^{2}\right]\rd\\
\propto&\int   \d \mathbf{M}  \exp \left[-\frac{\i}{2} \sum_{a,\mu}\left(M_{\mu}^{a}\right)^{2}\right]\ld\exp\left[-\i \sum_{i, a,\mu} \sqrt{\frac{\Lambda_{\mu}}{N}} \hat{\xi}_{i}^{\mu} y_{i}^{a} M_{\mu}^{a}\right]\rd,
\label{eq:Hubbard1}
\end{aligned}
\end{equation}
where $\d \mathbf{M}$ denotes $\left(\prod_{a,\mu} \d M^a_{\mu} \right)$. Using the fact that ${\xih}_i^\mu$ follows independently $\mathcal{N}(0,1)$, we complete the average in Eq.~(\ref{eq:Hubbard1}), leading to the following
result:
\begin{equation}
\begin{aligned}
\left\langle \exp \left[-\i\sum_{a,\mu, i} \sqrt{\frac{\Lambda_{\mu}}{N}} {M}^a_{\mu} y^a_i {\xih}^\mu_i\right]
\right\rangle &= \prod_{\mu, i}\exp\left[-\frac{1}{2}\frac{\Lambda_{\mu}}{N}\left(\sum_a {M_\mu^a y_i^a}\right)^2\right]\\
& = \prod_i \exp\left[-\frac{1}{2}\sum_{a,b} y_i^a y_i^b \frac{1}{N}\sum_{\mu}\Lambda_{\mu}{M}_\mu^a {M}_\mu^b \right].
\label{eq:average}
\end{aligned}
\end{equation}

Collecting Eq.~(\ref{eq:average}) and Eq.~(\ref{eq:Hubbard1}), we rewrite Eq.~(\ref{eq:Zn1}) as follows,
\begin{equation}
\begin{aligned}
\langle Z^n(\lambda) \rangle 
\propto&\int_{-\infty}^{\infty} \d \mathbf{Y} \d \mathbf{M}  \exp \left[-\frac{\i\lambda}{2}\sum_{i,a} (y_i^a)^2\right]\\
\times&
  \exp \left[-\frac{\i}{2} \sum_{\mu, a}\left(M_{\mu}^{a}\right)^{2}\right]\prod_i \exp\left[-\frac{1}{2}\sum_{a,b} y_i^a y_i^b \frac{1}{N}\sum_{\mu}\Lambda_{\mu}{M}_\mu^a {M}_\mu^b \right].
\label{eq:Zn2}
\end{aligned}
\end{equation}
By introducing the overlap $Q^{ab} \equiv \frac{1}{N}\sum_{i} y_i^a y_i^b$ together with
its conjugate variable $\hat{Q}^{ab}$ (introduced by applying the integral representation of the delta function enforcing the overlap definition), Eq.~(\ref{eq:Zn2}) can be further rewritten as
\begin{equation}
\begin{aligned}
\left\langle Z^n (\lambda)\right\rangle\propto &
\int \d\mathbf{Y}\d\mathbf{Q}\d\mathbf{\hat{Q}} \d\mathbf{{M}}\exp\left[ {\i}\sum_{a,b,i}y_i^a \hat{Q}^{ab}y_i^b-\frac{\i\lambda}{2}\sum_{a,i} (y_i^a)^2-\i N\sum_{a,b}\hat{Q}^{ab}Q^{ab}\right]\\
\times&\exp\left[-\frac{1}{2}\sum_{\mu}\sum_{a,b}\Lambda_{\mu} {M}_\mu^a {Q}^{ab} {M}_\mu^b-\frac{\i}{2}\sum_{a,\mu} {({M}_{\mu}^a)^2}\right],
\label{eq:Zn3}
\end{aligned}
\end{equation}
where $\d\mathbf{Q}$ is the short-hand of $\prod_{a,b}\d Q^{ab}$,
and $\d\mathbf{\hat{Q}}$ denotes $\prod_{a,b}\d \hat{Q}^{ab}$.
Notice that we can complete the integral $\int \d\mathbf{Y}$ by applying the multivariate Fresnel integral as follows
\begin{equation}
\int \prod_{i,a} \d y_i^a \exp\left[ \i\sum_{a,b,i}y_i^a \hat{Q}^{ab}y_i^b-\frac{\i\lambda}{2}\sum_{a,i} (y_i^a)^2\right] \propto [\mbox{det}(\lambda\mathbf{I}-2\mathbf{\hat{Q}})]^{-\frac{N}{2}},
\label{eq:int_y}
\end{equation}
where $\mathbf{\hat{Q}}$ is a $P\times P$ matrix with elements $\hat{Q}^{ab}$. 
In addition, we work out the integral over ${M}^a_\mu$:
\begin{equation}
\int \prod_{\mu, a}\d {M}^a_\mu \exp\left[-\frac{1}{2}\sum_{\mu}\sum_{a,b}{M}_\mu^a\Lambda_{\mu}{Q}^{ab}{M}_\mu^b-\frac{\i}{2}\sum_{a,\mu} (\hat{M}_{\mu}^a)^2\right] \propto \prod_{\mu=1}^P \left[ \mbox{det}(\mathbf{I}-\i\Lambda_{\mu}\mathbf{{Q}})\right]^{-\frac{1}{2}}.
\label{eq:int_m}
\end{equation}
Similarly, $\mathbf{{Q}}$ is the matrix with elements ${Q}^{ab}$.

Inserting Eq.~(\ref{eq:int_y}) and Eq.~(\ref{eq:int_m}) into Eq.~(\ref{eq:Zn3}), we rewrite Eq.~(\ref{eq:Zn3}) as
\begin{equation}
\begin{aligned}
\left\langle Z^n (\lambda)\right\rangle\propto &
\int \d\mathbf{Q}\d\mathbf{\hat{Q}} \exp\left[-Nf(\mathbf{Q},\mathbf{\hat{Q}})\right],
\label{eq:Zn4}
\end{aligned}
\end{equation}
where the free energy density $f$ is given by
\begin{equation}
f(\mathbf{Q}, \hat{\mathbf{Q}})=\mathrm{i} \operatorname{Tr} \hat{\mathbf{Q}}^{\mathrm{T}} \mathbf{Q}+\frac{1}{2 N} \sum_{\mu} \ln \operatorname{det}\left(\mathbf{I}-\mathrm{i} \Lambda_{\mu} \mathbf{Q}\right)+\frac{1}{2} \ln \operatorname{det}(\lambda \mathbf{I}-2 \hat{\mathbf{Q}}).
\label{eq:f}
\end{equation}
By applying the Laplace method in the thermodynamic limit (or we are interested in the asymptotic eigen-spectrum), we have
\begin{equation}
\ln \left\langle Z^n (\lambda)\right\rangle \approx \min_{\mathbf{Q},\mathbf{\hat{Q}}}\left[-Nf(\mathbf{Q},\mathbf{\hat{Q}})\right].
\end{equation}
Then the saddle-point equations can be obtained by minimizing $f$ with respect to $\mathbf{\hat{Q}}$ and $\mathbf{Q}$, i.e.,
$\frac{\partial f(\mathbf{Q},\mathbf{\hat{Q}})}{\partial \mathbf{\hat{Q}}}  = 0$ and $\frac{\partial f(\mathbf{Q},\mathbf{\hat{Q}})}{\partial \mathbf{Q}}  = 0$. The results are given by
\begin{subequations}
\begin{align}
&\hat{\mathbf{Q}}^{\mathrm{T}} \mathbf{Q}=\frac{1}{2} \lambda \mathbf{Q}+\frac{\mathrm{i}}{2} \mathbf{I},\label{sde1}\\
&-\mathrm{i} \lambda \mathbf{I}+\frac{1}{N} \sum_{\mu} \frac{\mathrm{i} \Lambda_{\mu}}{(\mathbf{I}-\mathrm{i} \Lambda_\mu \mathbf{Q})^{\mathrm{T}}}+(\mathbf{Q}^{\mathrm{T}})^{-1}=0.\label{sde2}
\end{align}
\end{subequations}
Note that $\hat{\mathbf{Q}}$ is now a function of $\mathbf{Q}$ [Eq.~(\ref{sde1})], $f(\mathbf{Q},\mathbf{\hat{Q}})$ can thus be simplified to $f(\mathbf{Q})$ as
\begin{equation}
f(\mathbf{Q})=\frac{\i}{2} \lambda \operatorname{Tr} \mathbf{Q}-\frac{n}{2}+\frac{1}{2 N} \sum_{\mu} \ln \operatorname{det}\left(\mathbf{I}-\mathrm{i} \Lambda_{\mu} \mathbf{Q}\right)-\frac{1}{2} \ln \operatorname{det}(\mathrm{i} \mathbf{Q}).
\label{eq:f2}
\end{equation}
Eq.~(\ref{sde2}) is the stationary condition of $f(\mathbf{Q})$.

To proceed, we make the replica symmetric assumption, i.e., ${Q}^{ab} = \delta_{ab} {Q}+(1-\delta_{ab}){q}$.
Under this assumption, $f(\mathbf{Q})$ can be written as a function of $Q$ and $q$:
\begin{equation}\label{fQq}
\begin{aligned} f(Q, q)=& \frac{\i n}{2} \lambda Q-\frac{n}{2}+\frac{1}{2 N} \sum_{\mu} \ln \left(1-\frac{\i n \Lambda_{\mu} q}{1-\i \Lambda_{\mu} Q+\i \Lambda_{\mu} q}\right)+\frac{n}{2 N} \sum_{\mu} \ln \left(1-\i \Lambda_{\mu} Q+\i \Lambda_{\mu} q\right) \\
&-\frac{1}{2} \ln \left(1+\frac{n q}{Q-q}\right)-\frac{1}{2} n \ln [\i(Q-q)].
\end{aligned}
\end{equation}
To derive Eq.~(\ref{fQq}), we have used the identity $\ln\operatorname{det}(\i\mathbf{Q})=\ln(1+\frac{nq}{Q-q})+n\ln[\i(Q-q)]$, and 
$\ln\operatorname{det}(\I-\i\Lambda_\mu\mathbf{Q})=\ln(1-\frac{\i n\Lambda_\mu q}{1-\i\Lambda_\mu(Q-q)})+n\ln(1-\i\Lambda_\mu(Q-q))$.
To derive the eigen-spectrum [see Eq.~(\ref{eq:rho_replica})], we first carry out the limit as follows,
\begin{equation}
\begin{aligned} F(Q, q)\equiv& \lim _{n \rightarrow 0} \frac{2 f(Q, q)}{n} \\=& \mathrm{i} \lambda Q-1-\frac{1}{N} \sum_{\mu} \frac{\mathrm{i} \Lambda_{\mu} q}{1-\mathrm{i} \Lambda_{\mu} Q+\mathrm{i} \Lambda_{\mu} q}+\frac{1}{N} \sum_{\mu} \ln \left(1-\mathrm{i} \Lambda_{\mu} Q+\mathrm{i} \Lambda_{\mu} q\right) \\ 
&-\frac{q}{Q-q}-\ln [\mathrm{i}(Q-q)].
\end{aligned}
\end{equation}
The function $F(Q,q)$ should be minimized with respect to $Q$ and $q$
according to the saddle-point condition, which gives:
\begin{align}
\label{eq:saddleq0}
&{q}=0,\\
\label{eq:saddleQ}
&\frac{1}{\i{Q}}-{\lambda}+\alpha \frac{1}{P} \sum_{\mu=1}^{P} \frac{1}{\Lambda_{\mu}^{-1}-\i Q}=0.
\end{align}
Comparing the definition of $F(Q,q)$ with Eq.~(\ref{eq:rho_replica}) and Eq.~(\ref{eq:rho}), we immediately arrive at
\begin{equation}
G_{\J} = \frac{\partial F(Q, q)}{\partial \lambda} = \i Q.
\end{equation}
Hence the Green function $G_{\J}$ must satisfy
\begin{equation}
\frac{1}{G_{\J}}-{\lambda}+\alpha \frac{1}{P} \sum_{\mu=1}^{P} \frac{1}{\Lambda_{\mu}^{-1}-G_{\J}}=0,
\label{eq:spetrum}
\end{equation}
where $\Lambda_\mu$ denotes the $\mu$-th diagonal element of the matrix $\bLam$. A solution of Eq.~(\ref{eq:spetrum}) can be used
to obtain the eigen-density at $\lambda$.

In summary, as $P\to\infty$, $\mu/P$ in the function $\Lambda_\mu$ [see Eq.~(\ref{eigX})] can be asymptotically mapped to the interval $[0,1]$. As a result,
$\Lambda_\mu$ turns out to be the function defined below,
\begin{equation}
 A(x) = c+2 \gamma \sum_{r=1}^{d} \cos \left(2 \pi rx\right),
 \end{equation}
 and $\frac{1}{P}\sum_{\mu=1}^{P} F(\Lambda_\mu)\simeq \int_{0}^1 dx F(A(x))$. We thus rewrite the spectrum
 equation as follows 
\begin{equation}
\frac{1}{G_{\mathbf{J}}}-\lambda+\alpha \int_0^1 dx \frac{A(x)}{1- G_{\mathbf{J}}A(x)}=0.
\label{eq:spetrum3}
\end{equation}
\subsection{Free probability method}
The free probability theory was developed to study the asymptotic spectral density of either sums or products of random matrices with special
symmetry properties~\cite{FP-2017}. For example, matrices $\mathbf{A}$ and $\mathbf{B}$ can be considered mutually free when their eigenvectors are almost 
surely orthogonal. We find that our current model satisfies this free property. More precisely, the coupling matrix can be decomposed into a sum of $P$ free matrices:
\begin{equation}
\mathbf{J} = \sum_{\mu=1}^P \mathbf{J}^{\mu},
\end{equation}
where $\mathbf{J}^{\mu} = \Lambda_{\mu}\frac{1}{N}\boldsymbol{{\xih}}^{\mu} (\boldsymbol{{\xih}}^{\mu})^{\t}$. 
$\boldsymbol{{\xih}}^{\mu}$ is the $\mu$th column of the matrix $\boldsymbol{{\xih}}$. 
By definition, $(\boldsymbol{{\xih}}^\mu)^{\t} \boldsymbol{{\xih}}^{\nu} = N\delta_{\mu\nu}$, and therefore we have:
\begin{equation}
\mathbf{J}^{\mu} \boldsymbol{{\xih}}^{\nu} = \Lambda_{\mu} \delta_{\mu \nu}\boldsymbol{{\xih}}^{\nu}.
\end{equation}
This shows that $\J^\mu$ is a rank-one matrix. It can be alternatively shown that $\tr(\J^\mu)=\Lambda_\mu\frac{1}{N}\tr\left(\boldsymbol{{\xih}}^{\mu} (\boldsymbol{{\xih}}^{\mu})^{\t}\right)=\Lambda_\mu$.
Thus the eigenvectors of $\mathbf{J}^{\mu}$ are exactly the patterns themselves. 
Furthermore, the eigenvectors of two different component matrices ($\J^{\mu}$) are almost surely orthogonal, 
which means that the component matrices are asymptotically free as $N\to\infty$.
According to the free probability theory, we then have
\begin{equation}
\mathcal{R}_{\J} =\sum_{\mu=1}^P \mathcal{\mathcal{R}_{\mathbf{J}^{\mu}}},
\label{eq:R_sum}
\end{equation}
where $\mathcal{R}$ denotes the R-transform, defined as
\begin{equation}
\mathcal{R}_{\J^\mu}(G_{\J^\mu}) \equiv \lambda(G_{\J^\mu}) - \frac{1}{G_{\J^\mu}},
\label{eq:R_transform}
\end{equation}
where $G_{\J^\mu}$ is the Green function defined before. 

Next we estimate the Stieltjes transform as follows,
\begin{equation}
	\begin{aligned}
	G_{\J^\mu}(\lambda) &=  \f{1}{N} \Tr(\lambda\I-\J^\mu )^{-1} 
	\\&= \f{1}{N} \sum_{k=0}^\infty \f{\Tr( \J^\mu )^k}{\lambda^{k+1}}
	\\& =  \f{1}{N} \l( \f{N}{\lambda} +\f{1}{\lambda} \sum_{k=1}^\infty \f{\Lambda_\mu^k}{\lambda^k} \r)
	\\& = \f{1}{N} \l( \f{N}{\lambda} -\f{1}{\lambda} +\f{1}{\lambda} \sum_{k=0}^\infty \f{\Lambda_\mu^k }{\lambda^k} \r)
	\\& =  \f{1}{N} \l( \f{N-1}{\lambda} +\f{1}{\lambda-\Lambda_\mu} \r)  \ .
\end{aligned}
\end{equation}
Note that we do not need to compute the disorder average here, as the Stieltjes transform of a rank-one matrix can be directly worked out.
More precisely, $\rho_{\J^\mu}(\lambda')=\frac{N-1}{N}\delta(\lambda')+\frac{1}{N}\delta(\lambda'-\Lambda_\mu)$,
and we can then estimate $G_{\J^\mu}(\lambda)=\langle(\lambda-\lambda')^{-1}\rangle_{\rho_{\J^\mu}}= \f{1}{N} \l( \f{N-1}{\lambda} +\f{1}{\lambda-\Lambda_\mu} \r)$.
However, to get the equation for $G_\J$, the property of the free matrix should be used, together with the R-transform, which we shall show below.

Then we calculate the functional inverse of the Stieltjes transform by
\begin{equation}
	G_{\bJ^\mu}=\f{1}{N} \l( \f{N-1}{\lambda(G_{\bJ^\mu})} +\f{1}{\lambda(G_{\bJ^\mu})-\Lambda_\mu} \r) \ ,
\end{equation}
and therefore
\begin{equation}\label{fptz}
	NG_{\bJ^\mu} \lambda^2-N(\Lambda_\mu G_{\bJ^\mu}+1)\lambda +(N-1)\Lambda_\mu =0  \ .
\end{equation}
Solving Eq.~(\ref{fptz}) of $\lambda (G)$, we have
\begin{equation}
	\begin{aligned}
	\lambda(G) &=  \f{N(\Lambda_\mu G+1)\pm\sqrt{ N^2(\Lambda_\mu G+1)^2-4NG(N-1)\Lambda_\mu    }}{2NG}
	\\& =  \f{N(\Lambda_\mu G+1)\pm\sqrt{ N^2(\Lambda_\mu G-1)^2+4NG\Lambda_\mu    }}{2NG}
	\\& = \f{N(\Lambda_\mu G+1)\pm\sqrt{ N^2(G\Lambda_\mu-1)^2+4NG\Lambda_\mu+\l( \f{2G\Lambda_\mu }{\Lambda_\mu G-1}\r)^2- \l( \f{2G\Lambda_\mu }{\Lambda_\mu G-1}\r)^2   }}{2NG}
	\\& \approx  \f{N(\Lambda_\mu G+1)\pm| N(\Lambda_\mu G-1)+ \f{2G\Lambda_\mu }{\Lambda_\mu G-1} |}{2NG} \ ,
\end{aligned} 
\end{equation}
where for large $N$, $\f{2G\Lambda_\mu }{\Lambda_\mu G-1}$ is negligible, and we neglect the subscript ($\J^\mu$)
for $G$. Using the property of the Stieltjes transform~\cite{Clean-2017}, i.e., $\lim_{|z|\to\infty}zG(z)=1$,
we can choose the correct root as follows
\begin{equation}
	\lambda(G_{\J^\mu}) = \f{1}{G_{\J^\mu}}+ \f{\Lambda_\mu }{N(1-\Lambda_\mu  G_{\J^\mu})} \ .
\end{equation}

By using the definition of the R-transform [Eq.~(\ref{eq:R_transform})],
we obtain the result:
\begin{equation}\label{rjmu}
	\mathcal{R}_{\J^\mu} (G_{\J^\mu}) = \f{\Lambda_\mu }{N(1-\Lambda_\mu G_{\J^\mu})} \ .
\end{equation}
Note that $G_{\J^\mu}$ is the argument of the $\mathcal{R}$ function. In fact, 
$\J^\mu/\Lambda_\mu$ forms a Wishart ensemble, whose R-transform $\mathcal{R}(z)=\frac{1}{N(1-z)}$
can be used to arrive at the same result [Eq.~(\ref{rjmu})].
According to the free sum [Eq.~(\ref{eq:R_sum})], we arrive at:
\begin{equation}
	\mathcal{R}_{\J} (G_{\J}) =\f{\a}{P}\sum_{\mu=1}^P \f{\Lambda_\mu }{1-\Lambda_\mu G_\J} =  \lambda(G_{\bJ})-\f{1}{G_{\bJ}}\ .
\end{equation}
Interestingly, the R-transform is related to the self-energy in physics~\cite{Zee-1996}; the latter is particularly useful in analyzing the asymptotic spectral properties of 
asymmetric random matrices (e.g., using the diagrammatic techniques). 
To conclude, the Stieltjes transform of the coupling matrix $\J$ is thus given by
\begin{equation}
	\f{1}{G_\J(\lambda)}-\lambda+ \f{\a}{P}\sum_{\mu=1}^P \f{\Lambda_\mu }{1-\Lambda_\mu G_\J(\lambda)} =0,
\end{equation}
which is the very equation we have derived using the replica method [see Eq.~(\ref{eq:spetrum})].
The eigenvalue distribution can be obtained by
\begin{equation}
	\p(\lam)=\frac{1}{\pi} \lim _{\epsilon \rightarrow 0^{+}} \operatorname{Im}  G_\bJ(\lambda-\i\e)\ .
\end{equation}

\section{Maximum eigenvalue}
To calculate the maximum eigenvalue of $\J$, which is related to the phase transition of paramagnetic phase to spin glass phase,
either the ground-state method or results from the eigenvalue spectrum (or vanishing imaginary-part method) can be used. In this section, we provide details about these two methods.
\subsection{Ground-state method}
Finding the maximum eigenvalue $\lambda_{\max}$ of $\mathbf{J}$ is equivalent 
to minimizing the following constrained optimization problem~\cite{Kabashima-2010}:
\begin{equation}
\lambda_{\max} = -\min_{\mathbf{u}^{\mathrm{T}}\mathbf{u} = N}\left\langle-\frac{1}{N}\mathbf{u}^{\mathrm{T}}\mathbf{J}\mathbf{u}\right\rangle,
\label{eq:max}
\end{equation}
where $\mathbf{u}$ is an arbitrary $N\times1$ vector with the $\ell_2$ length constraint $\|\mathbf{u}\|^2_2 =N$.
Eq.~(\ref{eq:max}) can be understood as follows, 
\begin{equation}
-\frac{1}{N}\mathbf{u}^{\mathrm{T}}\mathbf{J}\mathbf{u} = -\frac{1}{N}\sum_i  \lambda_i u_i^2\ge - \lambda_{\max},
\end{equation}
where we define $\lambda_i$ as the $i^{\mathrm{th}} $ eigenvalue of $\mathbf{J}$, and an orthogonal transformation of $\mathbf{u}$ preserves its length.

If we consider $\mathcal{H}(\mathbf{u})\equiv-\mathbf{u}^{\mathrm{T}}\mathbf{J}\mathbf{u}$ as the Hamiltonian of an interacting-particle system at a state $\mathbf{u}$,
then the optimal $\mathbf{u}$ is exactly the ground state of the system. The task of finding the maximum eigenvalue can be further interpreted as finding the ground state of system with $\mathcal{H}(\mathbf{u})$.
The canonical partition function of the system reads:
\begin{equation}
\mathcal{Z}(\beta) = \int \d\mathbf{u}\;\delta(\mathbf{u}^{\mathrm{T}}\mathbf{u}-N)\; \exp(\beta \mathbf{u}^{\mathrm{T}}\mathbf{J}\mathbf{u}),
\end{equation}
where $\beta$ is the inverse temperature, and $\d \mathbf{u}\equiv\prod_{i}\d u_i$. The delta function enforces
the length constraint. 
It then follows that
$\lambda_{\max} = \frac{1}{N}\lim_{\beta\to\infty} \frac{1}{\beta} \langle\ln\mathcal{Z}(\beta)\rangle$. 
In the following, we apply the replica trick to perform the quenched average:
\begin{equation}
\lambda_{\max} = \frac{1}{N}\lim_{\beta\to\infty} \frac{1}{\beta}\lim_{n\to 0}\frac{\ln\langle\mathcal{Z}^n(\beta)\rangle}{n},
\label{eq:lbd_max}
\end{equation}
where
\begin{equation}
\left\langle \mathcal{Z}^{n}(\beta)\right\rangle=\int_{-\infty}^{\infty} \prod_{a=1}^{n} \left[\d \mathbf{u}^{a} \delta\left(\left(\mathbf{u}^{a}\right)^{\mathrm{T}} \mathbf{u}^{a}-N\right)\right]\left\langle \exp \left(\frac{\beta}{N} \sum_{a=1}^{n}\left(\mathbf{u}^{a}\right)^{\mathrm{T}} {\boldsymbol{{\xih}}}^{\mathrm{T}} \mathbf{\Lambda} {\boldsymbol{{\xih}}} \mathbf{u}^{a}\right)\right\rangle.
\end{equation}
Hereafter, $\mathbf{u}^{a}$ denotes the $a^{\mbox{th}}$ replica of $\mathbf{u}$. The disorder average refers to the average over the transformed patterns $\hat{\boldsymbol{\xi}}$.

Using the Hubbard-Stratonovich transform to linearize the quadratic term of ${\xih}_i^{\mu} u_i^a$, we have
\begin{equation}
\begin{aligned}
\ld\mathcal{Z}^{n}(\beta)\rd\propto& \int \d \boldsymbol{\M}  \d \boldsymbol{U}  \prod_{a=1}^{n} \delta\left(\left(\mathbf{u}^{a}\right)^{\mathrm{T}} \mathbf{u}^{a}-N\right) \\
& \exp \left[-\sum_{\mu, a}\left(\M_{\mu}^{a}\right)^{2}\right]\left\langle\exp\left[-2 \sqrt{\frac{\beta}{N}} \sum_{a, \mu, i} \sqrt{\Lambda_{\mu}} \M_{\mu}^{a} {{\xih}}_{i}^{\mu} u_{i}^{a}\right]\right\rangle,
\label{eq:Zn_beta1}
\end{aligned}
\end{equation}
where $\d \boldsymbol{\M}\equiv\prod_{a,\mu} \d \M_{\mu}^a$, and $\d \boldsymbol{U} \equiv \prod_{i,a} \d u_i^a $.
$\langle.\rangle$ is the easy-to-calculate disordered average which is given by
\begin{equation}
\begin{aligned}
\left\langle\exp \left[-2 \sqrt{\frac{\beta}{N}} \sum_{a, \mu, i} \sqrt{\Lambda_{\mu}} \M_{\mu}^{a} \xih_{i}^{\mu} u_{i}^{a}\right]\right\rangle 
=\exp \left[\frac{2\beta}{N} \sum_{a, b, \mu, i} \Lambda_{\mu} \M_{\mu}^{a} \M_{\mu}^{b} u_{i}^{a} u_{i}^{b}\right].
\label{eq:average2}
\end{aligned}
\end{equation}
By introducing the overlap $\Q^{ab}\equiv\frac{1}{N}\sum_i u_i^a u_i^b$ ($\Q^{aa} = 1$ 
because of the constraint on $\mathbf{u}$) accompanied by its conjugate variable $\hat{\Q}^{ab}$, we recast Eq.~(\ref{eq:Zn_beta1}) as
\begin{equation}
\begin{aligned}
\ld\mathcal{Z}^{n}(\beta)\rd \propto 
\int & \d \boldsymbol{U} \d \boldsymbol{\M} \d \boldsymbol{\Q} \d \boldsymbol{\hat{\Q}}\; \\
\exp \left[-\sum_{a, b, i} u_{i}^{a} \hat{\Q}^{a b} u_{i}^{b} \right]\exp\left[N\sum_{a,b} \Q^{ab} \hat{\Q}^{ab}\right] 
\times&\exp \left[-\sum_{\mu, a} \left({\M}_{\mu}^a\right)^2+2 \beta \sum_{\mu,a,b} \Lambda_{\mu} {\M}_{\mu}^{a} \Q^{ab} {\M}_{\mu}^{b}\right],
\end{aligned} 
\end{equation}
where $\d \boldsymbol{\Q} \equiv \prod_{a<b} \d\Q^{ab}$ and $\d \boldsymbol{\hat{\Q}} \equiv \prod_{a<b} \d\hat{\Q}^{ab}$ (note that the replica matrix is symmetric). 
After completing the integral over $\boldsymbol{\M}$ and $\boldsymbol{U}$, we obtain
\begin{equation}
\begin{aligned}
\ld\mathcal{Z}^{n}(\beta)\rd \propto 
\int \d \boldsymbol{\Q} \d \boldsymbol{\hat{\Q}}
\;&\exp \left[-\frac{N}{2} \ln \operatorname{det} \hat{\boldsymbol{\Q}}\right]\exp\left[N \operatorname{Tr} \hat{\boldsymbol{\Q}}^{\t} \boldsymbol{\Q}\right] \\
\times&\exp \left[-\frac{1}{2} \sum_{\mu=1}^{P} \ln \operatorname{det}\left(\mathbf{I}-2 \beta \Lambda_{\mu} \boldsymbol{\Q}\right)\right].
\label{eq:Zn_beta2}
\end{aligned} 
\end{equation}
Equation~($\ref{eq:Zn_beta2}$) can be reorganized into a concise form as
\begin{equation}
\left\langle\mathcal{Z}^{n}(\beta) \right\rangle\propto \int \d \boldsymbol{\Q} \d \hat{\boldsymbol{\Q}} \;\exp\left(-N \mathcal{F}(\boldsymbol{\Q},\hat{\boldsymbol{\Q}}) \right),
\end{equation}
with the free energy density given by
\begin{equation}
\mathcal{F}(\boldsymbol{\Q},\hat{\boldsymbol{\Q}}) = -\operatorname{Tr} \hat{\boldsymbol{\Q}}^{\mathrm{T}} \boldsymbol{\Q}+\frac{1}{2} \ln \operatorname{det} \hat{\boldsymbol{\Q}}+\frac{1}{2 N} \sum_{\mu=1}^{P} \ln \operatorname{det}\left(\mathbf{I}-2 \beta \Lambda_{\mu} \boldsymbol{\Q}\right).
\end{equation}

In the thermodynamic limit, the saddle point approximation requires that
$\frac{\partial \mathcal{F}(\boldsymbol{\Q},\hat{\boldsymbol{\Q}})}{\partial\hat{\boldsymbol{\Q}}}=0$, 
which leads to
\begin{equation}
(2 \boldsymbol{\Q})^{-1}=\hat{\boldsymbol{\Q}}^{\mathrm{T}}.
\label{eq:saddle2}
\end{equation}
Under the replica symmetric ansatz, i.e., $\Q^{ab}=\delta_{ab} +\q(1-\delta_{ab})$, inserting Eq.~(\ref{eq:saddle2}) 
into $\mathcal{F}(\boldsymbol{\Q},\hat{\boldsymbol{\Q}})$, we have:
\begin{equation}
\begin{aligned}
\mathcal{F}(\q)=& -\frac{n}{2}-\frac{1}{2} \ln \left(1+\frac{n \q}{1-\q}\right)-\frac{n}{2} \ln [2(1-\q)] \\
&+\frac{1}{2 N} \sum_{\mu=1}^{P} \ln \left[1+\frac{-2 n \beta \Lambda_{\mu} \q}{1-2 \beta \Lambda_{\mu}(1-\q)}\right]+\frac{n}{2 N} \sum_{\mu=1}^{P} \ln \left[1-2 \beta \Lambda_{\mu}(1-\q)\right].
\end{aligned}
\end{equation}
We then take the replica limit $n\to0$ as follows,
\begin{equation}
\begin{aligned}
\lim_{n\to 0} \frac{-\mathcal{F}(\q)}{n} &=
\frac{1}{2}+\frac{1}{2} \frac{\q}{1-\q}+\frac{1}{2} \ln [2(1-\q)] \\
&+\frac{1}{2 N} \sum_{\mu=1}^{P} \frac{2 \beta \Lambda_{\mu} \q}{1-2 \beta \Lambda_{\mu}(1-\q)}-\frac{1}{2 N} \sum_{\mu=1}^{P} \ln \left[1-2 \beta \Lambda_{\mu}(1-\q)\right]\\
&\equiv g(\q).
\label{eq:g}
\end{aligned}
\end{equation}
It then requires that $\frac{\partial g(\q)}{\partial \q}=0$, thereby leading to
\begin{equation}
\frac{1}{4 \beta^{2}(1-\q)^{2}}=\frac{1}{N} \sum_{\mu=1}^{P} \frac{\Lambda_{\mu}^{2}}{\left[1-2 \Lambda_{\mu} \beta(1-\q)\right]^{2}}.
\label{eq:saddle3}
\end{equation}

It is physically intuitive that $\q\to1$ as $\beta\to \infty$. We then assume that the value of $\beta(1-\q)$ in the zero temperature limit is
finite, being the fixed point solution of Eq.~(\ref{eq:saddle3}). 
 We then define $C\equiv2\beta(1-\q)$, 
 and comparing Eq.~(\ref{eq:g}) with Eq.~(\ref{eq:lbd_max}), we immediately arrive at
 \begin{equation}
 \begin{aligned}
\lambda_{\max }=&\lim _{\beta \rightarrow \infty} \frac{1}{\beta}\; g(\q)\\
=& \lim _{\beta \rightarrow \infty}\left\{\frac{1}{2 \beta}+\frac{1}{2} \frac{\q}{\beta(1-\q)}+\frac{1}{2 \beta} \ln [2(1-\q)]\right.\\
&\left.+\frac{1}{2 N} \sum_{\mu} \frac{2 \Lambda_{\mu} \q}{1-2 \beta \Lambda_{\mu}(1-\q)}-\frac{1}{2 \beta N} \sum_{\mu} \ln \left[1-2 \beta \Lambda_{\mu}(1-\q)\right]\right\}\\
=&
\frac{1}{C}+\frac{1}{N} \sum_{\mu=1}^{P} \frac{\Lambda_{\mu}}{1-\Lambda_{\mu} C}.
\label{eq:lbd_max1}
\end{aligned}
\end{equation}
Equation~(\ref{eq:saddle3}) becomes
\begin{equation}
\frac{1}{C^{2}}=\frac{1}{N} \sum_{\mu=1}^{P} \frac{\Lambda_{\mu}^{2}}{\left(1-\Lambda_{\mu} C\right)^{2}}.
\label{eq:c}
\end{equation}
After a solution of $C$ is obtained from Eq.~(\ref{eq:c}), the maximum eigenvalue can be obtained according to Eq.~(\ref{eq:lbd_max1}).
Note that starting from a relatively small value of $C$, $\lambda_{\rm min}$ where the spectral density also vanishes
could be reached.
As explained before,
the maximum-eigenvalue equation can be equivalently written into the following form,
\begin{equation}
\frac{1}{C^{2}}=\alpha \int_0^1\mathrm{d}x \frac{A^{2}(x)}{\left(1- CA(x)\right)^{2}}.
\label{eq:maxeigenvalue2}
\end{equation}

\subsection{Vanishing-imaginary-part method}
A necessary condition of the maximum eigenvalue is that the spectral density vanishes.
Setting $G_{\mathbf{J}} = X + \i Y$, where $X$ and $Y$ are both real-valued, we rewrite Eq.~(\ref{eq:spetrum}) as
\begin{equation}
\lambda =
\frac{\alpha}{P} \sum_{\mu=1}^P\frac{\Lambda_{\mu}(1-X \Lambda_{\mu}+\i \Lambda_{\mu} Y)}{(1-\Lambda_{\mu} X)^{2}+\Lambda_{\mu}^{2} Y^{2}}+\frac{X-\i Y}{X^{2}+Y^{2}}.
\label{eq:spectrum2_IR}
\end{equation}
Now we divide Eq.~(\ref{eq:spectrum2_IR}) into real and imaginary parts as follows
\begin{align}
\label{eq:spectrum2_R}
\lambda&=\frac{\alpha}{P}  \sum_{\mu=1}^P \frac{\Lambda_{\mu}}{(1-\Lambda_{\mu} X)}+\frac{1}{X},\\
\label{eq:spectrum2_I}
\frac{Y}{X^{2}+Y^{2}}&=\frac{\alpha}{P} \sum_{\mu=1}^P \frac{\Lambda^{2}_{\mu} Y}{(1-\Lambda_{\mu} X)^{2}+\Lambda_{\mu}^{2} Y^{2}}.
\end{align}
It is clear that $Y=0$ is always a solution, corresponding to $\rho(\lambda)=0$. Equation~(\ref{eq:spectrum2_I}) in the $Y\to0$ limit reduces to
\begin{equation}
\frac{1}{X^{2}}=\frac{\alpha}{P} \sum_{\mu=1}^P \frac{\Lambda^{2}_{\mu} }{(1-\Lambda_{\mu} X)^{2}},
\end{equation} 
which is exactly identical to Eq.~(\ref{eq:c}) after replacing $X$ by $C$. We finally remark that the vanishing-imaginary-part method applies to
the case of continuous spectral density.

\section{Relationship between the maximum eigenvalue and phase transition}
The Hamiltonian of the system [see Eq.~(\ref{eq2})] can be recast into the following form,
\begin{equation}
\begin{aligned}
\mathcal{H}(\mathbf{s)} &= -\frac{1}{2}\sum_{ij}J_{ij} s_i s_j\\
& = -\frac{1}{2N} \sum_{ij}\sum_{\rho\mu} \xi_i^\rho X_{\rho\mu} \xi_j^{\mu} s_i s_j\\
& = -\frac{1}{2N} \sum_{ij}\sum_{\rho\mu} \xi_i^\rho \xi_j^{\mu} s_i s_j \sum_{\sigma}\Lambda_{\sigma} O_{\sigma}^{\mu} O_{\sigma}^{\rho} \\
& = -\frac{1}{2N} \sum_{\sigma}  \left(\sqrt{\Lambda_{\sigma}}\sum_{i,\mu}\xi_i^{\mu}O_\sigma^\mu s_i\right)^2,
\end{aligned}
\end{equation}
where $s_i$ denotes the binary state of the neuron $i$, taking value of $+1$ or $-1$, $\mathbf{s}$ is the 
state configuration of all neurons,
and we have used the spectral decomposition of the circulant matrix $\X$. 
Then the partition function reads:
\begin{equation}
\begin{aligned}
Z &= \sum_{\mathbf{s}} \exp(-\beta \mathcal{H}(\mathbf{s}))\\
& = \sum_{\mathbf{s}} \prod_{\mu=1}^{P} \int \frac{\mathrm{d} z_{\mu}}{\sqrt{2\pi/\beta}}\exp\left(-\frac{\beta}{2}z_\mu^2+\frac{\beta}{\sqrt{N}}\sum_{i=1}^N z_\mu \Xi_i^\mu s_i\right),
\end{aligned}
\end{equation}
where we have used the Hubbard-Stratonovich transformation, 
and $\Xi_{i}^{\mu} \equiv \sqrt{\Lambda_{\mu}}\sum_{\rho=1}^{P} O_{\mu}^{\rho} \xi_{i}^{\rho}$. 
This expression shows that our associative memory network can be mapped to a 
restrict Boltzman machine (RBM), where $N$ binary visible nodes $s_i$ and $P$ 
continuous (Gaussian) hidden nodes $z_{\mu}$ interact through couplings $\Xi_{i}^{\mu}$. 
By the central-limit theorem, as $N\to\infty$, $\Xi_i^{\mu}$ is an i.i.d random 
variable and obeys $\mathcal{N}(0,\Lambda_{\mu})$ where $\Lambda_\mu\geq0$. 

Next, we derive the Thouless-Anderson-Palmer equations (TAP) from 
belief propagation (BP) equations. The BP equations can be written 
in a standard procedure once the factor graph is given. We will follow the strategy 
detailed in Ref.~\cite{Mezard-2017}.

We first define $m_{i\to\mu}(s_i)$ as the message sending from 
node $i$ to node $\mu$, and $m_{\mu\to i}(z_{\mu})$ as the message 
sending from $\mu$ to $i$. A physical interpretation of these messages
is that $m_{i\to\mu}(s_i)$ is the cavity distribution of $s_i$ in the 
absence of node $\mu$ and $m_{\mu\to i}(z_\mu)$ is the cavity distribution 
of $z_\mu$ in the absence of node $i$~\cite{Huang-2017}. 

Under the Bethe approximation~\cite{MM-2009}, i.e., the correlation between 
two nodes decays fast with the distance separating them, we have
\begin{align}
&m_{i \rightarrow \mu}\left(s_{i}\right) \propto \prod_{\nu \neq \mu} \hat{m}_{\nu \rightarrow i}\left(s_{i}\right), \\
&m_{\mu \rightarrow i}\left(z_{\mu}\right) \propto e^{-\frac{\beta}{2} z_{\mu}^{2}} \prod_{j \neq i} \hat{m}_{j \rightarrow \mu}\left(z_{\mu}\right),
\end{align}
where the two auxiliary messages are defined as
\begin{align}
&\hat{m}_{\mu \rightarrow i}\left(s_{i}\right)=\int \frac{d z_{\mu}}{\sqrt{2 \pi / \beta}} m_{\mu \rightarrow i}\left(z_{\mu}\right) \exp \left(\frac{\beta}{\sqrt{N}} \Xi_{i}^{\mu} s_{i} z_{\mu}\right), \\
&\hat{m}_{i \rightarrow \mu}\left(z_{\mu}\right)=\sum_{s_{i}} m_{i \rightarrow \mu}\left(s_{i}\right) \exp \left(\frac{\beta}{\sqrt{N}} \Xi_{i}^{\mu} s_{i} z_{\mu}\right).
\end{align}
Because $s_i$ is a binary variable, we can parameterize the messages involving $s_i$ as follows
\begin{align}
&m_{i \rightarrow \mu}\left(s_{i}\right) \propto e^{\beta h_{i \rightarrow \mu} s_{i}}, \\
&\hat{m}_{\mu \rightarrow i}\left(s_{i}\right) \propto e^{\beta \hat{h}_{\mu \rightarrow i} s_{i}},
\end{align}
where $h_{i\to\mu}$ is the cavity field, and $\hat{h}_{\mu \rightarrow i}$ is the conjugate cavity field.
It then follows that the BP equations are given by
\begin{align}
\label{eq:h_sum}
&h_{i \rightarrow \mu}=\sum_{\nu \neq \mu} \hat{h}_{\nu \rightarrow i}, \\
\label{eq:hhat}
&e^{\beta \hat{h}_{\mu \rightarrow i} s_{i}} \propto \int d z_{\mu} m_{\mu \rightarrow i}\left(z_{\mu}\right) \exp \left(\frac{\beta}{\sqrt{N}} \Xi_{i}^{\mu} s_{i} z_{\mu}\right), \\
\label{eq:mhat}
&\hat{m}_{i \rightarrow \mu}\left(z_{\mu}\right)=\frac{1}{2\cosh(\beta h_{i\to\mu})} \sum_{s_{i}} \exp \left(\beta h_{i \rightarrow \mu} s_{i}+\frac{\beta}{\sqrt{N}} \Xi_{i}^{\mu} s_{i} z_{\mu}\right)=\frac{\cosh \left(\beta h_{i \rightarrow \mu}+\frac{\beta}{\sqrt{N}} \Xi_{i}^{\mu} z_{\mu}\right)}{\cosh(\beta h_{i\to\mu})}, \\
\label{eq:m_prod}
&m_{\mu \rightarrow i}\left(z_{\mu}\right) \propto e^{-\frac{\beta z_{\mu}^2}{2}} \prod_{j \neq i} \hat{m}_{j \rightarrow \mu}\left(z_{\mu}\right).
\end{align}

We further assume that $m_{\mu\to i}(z_{\mu})$ is a Gaussian distribution with mean $a_{\mu\to i}$ 
and variance $c_{\mu\to i}$, which holds in the phase where there is no condensation 
on any specific pattern. In this case, we can expand Eq.~(\ref{eq:mhat}) in the $N\to\infty$ limit:
\begin{equation}
\hat{m}_{i \rightarrow \mu}\left(z_{\mu}\right)= \exp \left\{\beta \frac{\Xi_{i}^{\mu}}{\sqrt{N}} z_{\mu} \tanh \left(\beta h_{i \rightarrow \mu}\right)
+\frac{\beta^{2}\left(\Xi_{i}^{\mu}\right)^{2}}{2 N} z_{\mu}^{2}\left[1-\tanh ^{2}\left(\beta h_{i \rightarrow \mu}\right)\right]\right\}.
\end{equation}
Inserting this Taylor expansion into Eq.~(\ref{eq:m_prod}), we have
\begin{align}
&c_{\mu \rightarrow i} =\frac{1}{\beta} \frac{1}{1-(\beta / N) \sum_{j(\neq i)}\left(\Xi_{j}^{\mu}\right)^{2}\left[1-\tanh ^{2}\left(\beta h_{j \rightarrow \mu}\right)\right]} \\
&a_{\mu \rightarrow i} =\frac{1}{\sqrt{N}} \frac{\sum_{j(\neq i)} \Xi_{j}^{\mu} \tanh \left(\beta h_{j \rightarrow \mu}\right)}{1-(\beta / N) \sum_{j(\neq i)}\left(\Xi_{j}^{\mu}\right)^{2}\left[1-\tanh ^{2}\left(\beta h_{j \rightarrow \mu}\right)\right]}.
\end{align}
With the Gaussian form of $m_{\mu\to i}(z_\mu)$,  Eq.~(\ref{eq:h_sum}) and Eq.~(\ref{eq:hhat}) can be simplified as
\begin{equation}
h_{i \rightarrow \mu}=\sum_{\nu \neq \mu} \frac{\Xi_{i}^{\nu}}{\sqrt{N}} a_{\nu \rightarrow i},
\end{equation}
which implies that the full local field is given by
\begin{equation}
H_{i} =\sum_{\nu} \frac{\Xi_{i}^{\nu}}{\sqrt{N}} a_{\nu\rightarrow i},
\end{equation}
and
\begin{equation}
\label{eq:A_mu}
A_{\mu} =\frac{1}{\sqrt{N}} \frac{\sum_{j} \Xi_{j}^{\mu} \tanh \left(\beta h_{j \rightarrow \mu}\right)}{1-(\beta / N) \sum_{j}\left(\Xi_{j}^{\mu}\right)^{2}\left[1-\tanh ^{2}\left(\beta h_{j \rightarrow \mu}\right)\right]}.
\end{equation}

It is worth noting that
\begin{equation}
h_{j \rightarrow \mu}=H_{j}-\frac{\Xi_{j}^{\mu}}{\sqrt{N}} a_{\mu \rightarrow j},
\label{eq:h_H}
\end{equation}
which implies that $h_{j \rightarrow \mu} \approx H_{j}$ in the large $N$ limit. Hence, we have
\begin{equation}
\frac{1}{N} \sum_{j}\left(\Xi_{j}^{\mu}\right)^{2}\left[1-\tanh ^{2}\left(\beta h_{j \rightarrow \mu}\right)\right] \approx \frac{1}{N} \sum_{j}\left(\Xi_{j}^{\mu}\right)^{2}\left[1-\tanh ^{2}\left(\beta H_{j}\right)\right].
\end{equation}
Then, Eq.~(\ref{eq:A_mu}) becomes
\begin{equation}
A_{\mu}=\frac{1}{\sqrt{N}} \frac{\sum_{j} \Xi_{j}^{\mu} \tanh \left(\beta h_{j \rightarrow \mu}\right)}{1-\frac{\beta}{N} \sum_{j}\left(\Xi_{j}^{\mu}\right)^{2}\left[1-\tanh ^{2}\left(\beta H_{j}\right)\right]}.
\end{equation}
Along the similar line,
\begin{equation}
a_{\mu \rightarrow i} = A_{\mu}-\frac{1}{1-\frac{\beta}{N} \sum_{j}\left(\Xi_{j}^{\mu}\right)^{2}\left[1-\tanh ^{2}\left(\beta H_{j}\right)\right]} \frac{1}{\sqrt{N}} \Xi_{i}^{\mu} \tanh \left(\beta h_{i \rightarrow \mu}\right).
\label{eq:a_mu_i}
\end{equation}
By replacing $h_{i\to\mu}$ with $H_i$ in the second term of Eq.~(\ref{eq:a_mu_i}) 
(this approximation becomes exact in the large-$N$ limit), we have
\begin{equation}
a_{\mu \rightarrow i} \simeq A_{\mu}-\frac{1}{1-\frac{\beta}{N} \sum_{j}\left(\Xi_{j}^{\mu}\right)^{2}\left[1-\tanh ^{2}\left(\beta H_{j}\right)\right]} \frac{1}{\sqrt{N}} \Xi_{i}^{\mu} \tanh \left(\beta H_{i }\right).
\end{equation}
Finally, we derive $H_i$ without any cavity quantities in the following formula:
\begin{equation}
\begin{aligned}
H_{i} &=\sum_{\mu} \frac{\Xi_{i}^{\mu}}{\sqrt{N}} a_{\mu \rightarrow i} \\
&=\sum_{\mu} \frac{\Xi_{i}^{\mu}}{\sqrt{N}} A_{\mu}-\alpha \frac{1}{P}\left\{\sum_{\mu} \frac{(\Xi_{i}^{\mu})^2}{1-\frac{\beta}{N} \sum_{j}\left(\Xi_{j}^{\mu}\right)^{2}\left[1-\tanh ^{2}\left(\beta H_{j}\right)\right]}\right\} \tanh \left(\beta H_{i}\right).
\label{eq:H_expand}
\end{aligned}
\end{equation}

We next derive the self-consistent equation for $A_{\mu}$ as follows,
\begin{equation}
\begin{aligned} 
A_{\mu} &=\frac{1}{\sqrt{N}} \frac{\sum_{j} \Xi_{j}^{\mu} \tanh \left(\beta\left(H_{j}-\frac{\Xi_{j}^{\mu}}{\sqrt{N}} a_{\mu \rightarrow j}\right)\right)}{1-\frac{\beta}{N} \sum_{j}\left(\Xi_{j}^{\mu}\right)^{2}\left[1-\tanh ^{2}\left(\beta H_{j}\right)\right]} \\ 
&=\frac{1}{\sqrt{N}} \frac{\sum_{j} \Xi_{j}^{\mu} \tanh \left(\beta\left(H_{j}-\frac{\Xi_{j}^{\mu}}{\sqrt{N}} A_{\mu}\right)\right)}{1-\frac{\beta}{N} \sum_{j}\left(\Xi_{j}^{\mu}\right)^{2}\left[1-\tanh ^{2}\left(\beta H_{j}\right)\right]} \\ 
&=\frac{1}{1-\frac{\beta}{N} \sum_{j}\left(\Xi_{j}^{\mu}\right)^{2}\left[1-\tanh ^{2}\left(\beta H_{j}\right)\right]} \sum_{j} \frac{\Xi_{j}^{\mu}}{\sqrt{N}}\left[\tanh \left(\beta H_{j}\right)-\beta \frac{\Xi_{j}^{\mu}}{\sqrt{N}}\left(1-\tanh ^{2}\left(\beta H_{j}\right)\right) A_{\mu}\right] .
\label{eq:A_expand}
\end{aligned}
\end{equation}
Solving Eq.~(\ref{eq:A_expand}) for $A_\mu$ gives
\begin{equation}
A_{\mu}=\frac{1}{\sqrt{N}} \sum_{j} \Xi_{j}^{\mu} \tanh \left(\beta H_{j}\right).
\label{eq:A_simple}
\end{equation}

Inserting Eq.~(\ref{eq:A_simple}) into Eq.~(\ref{eq:H_expand}), we have:
\begin{equation}
H_{i}=\sum_{\mu} \frac{\Xi_{i}^{\mu}}{\sqrt{N}} \frac{1}{\sqrt{N}} \sum_{j} \Xi_{j}^{\mu} \tanh \left(\beta H_{j}\right)-\alpha \frac{1}{P}\left\{\sum_{\mu} \frac{\left(\Xi_{i}^{\mu}\right)^{2}}{1-\frac{\beta}{N} \sum_{j}\left(\Xi_{j}^{\mu}\right)^{2}\left[1-\tanh ^{2}\left(\beta H_{j}\right)\right]}\right\} \tanh \left(\beta H_{i}\right).
\label{eq:last_to_TAP}
\end{equation}
Then, the magnetization reads
\begin{equation}
m_{i}=\tanh \left(\beta \sum_{j} J_{i j} m_{j}-\alpha \beta \frac{1}{P}\left\{\sum_{\mu} \frac{(\Xi_i^\mu)^2}{1-\frac{\beta}{N} \sum_{j}\left(\Xi_{j}^{\mu}\right)^{2}\left[1-m_j^2\right]}\right\} m_{i}\right).
\end{equation}
As $N\to \infty$, we have
\begin{equation}
m_{i}=\tanh \left(\beta \sum_{j} J_{i j} m_{j}-\alpha \beta \frac{1}{P}\left\{\sum_{\mu} \frac{\Lambda_{\mu}}{1-{\beta} \Lambda_{\mu}\left(1-q\right)}\right\} m_{i}\right),
\label{eq:tap}
\end{equation}
where $q=\frac{1}{N}\sum_i^N m_i^2$ is the spin glass order parameter of the model.

Lastly, we determine the critical temperature for the transition 
from the paramagnetic phase ($q=0$) to the spin-glass phase ($q>0$). 
By expanding Eq.~(\ref{eq:tap}) around $m_{i}=0$ and $q=0$, we have
\begin{equation}
m_{i}=\beta \sum_{j} J_{i j} m_{j}-\alpha \beta \frac{1}{P}\left\{\sum_{\mu} \frac{\Lambda_{\mu}}{1-{\beta} \Lambda_{\mu}}\right\} m_{i}.
\label{eq:tap_expand}
\end{equation}
According to the linear stability analysis, the solution of $\bm=0$ to Eq.~(\ref{eq:tap_expand}) is stable only when
\begin{equation}
\beta \lambda_{\max} - \alpha \beta \frac{1}{P}\left\{\sum_{\mu} \frac{\Lambda_{\mu}}{1-{\beta} \Lambda_{\mu}}\right\}<1.
\end{equation}
Hence, the critical $\beta_g$ satisfies the following equation
\begin{equation}
\beta_g \lambda_{\max} - \alpha \beta_g \frac{1}{P}\left\{\sum_{\mu} \frac{\Lambda_{\mu}}{1-{\beta_g} \Lambda_{\mu}}\right\}=1.
\label{eq:beta_g}
\end{equation}
We find that Eq.~(\ref{eq:beta_g}) agrees with the result obtained from the 
replica computation~\cite{Huang-2021}. The replica computation gives
\begin{equation}\label{repres}
\alpha\frac{1}{P}\left\{\sum_{\mu} \frac{\Lambda_{\mu}^2}{(T_g- \Lambda_{\mu})^2}\right\}=1.
\end{equation}
By inserting the the expression of $\lambda_{\max}$ [see Eq.~(\ref{eq:lbd_max1})] into Eq.~(\ref{eq:beta_g}), we have
\begin{equation}
\beta_g \frac{1}{C}+\alpha\beta_g\frac{1}{P} \sum_{\mu} \frac{\Lambda_{\mu}}{1-\Lambda_{\mu} C} - \alpha \beta_g \frac{1}{P}\left\{\sum_{\mu} \frac{\Lambda_{\mu}}{1-{\beta_g} \Lambda_{\mu}}\right\}=1.
\label{eq:beta_g1}
\end{equation}
It is clear that $C=\beta_g$ is a solution of Eq.~(\ref{eq:beta_g1}).
The derivation of the maximum eigenvalue [see Eq.~(\ref{eq:c})] implies that
\begin{equation}
\frac{1}{C^{2}}=\frac{1}{N} \sum_{\mu=1}^{P} \frac{\Lambda_{\mu}^{2}}{\left(1-\Lambda_{\mu} C\right)^{2}}.
\end{equation}
Substituting $C=\beta_g$ recovers the result of the replica computation, i.e., Eq.~(\ref{repres}).

\begin{figure}
     \includegraphics[bb=2 5 848 525,width=\textwidth]{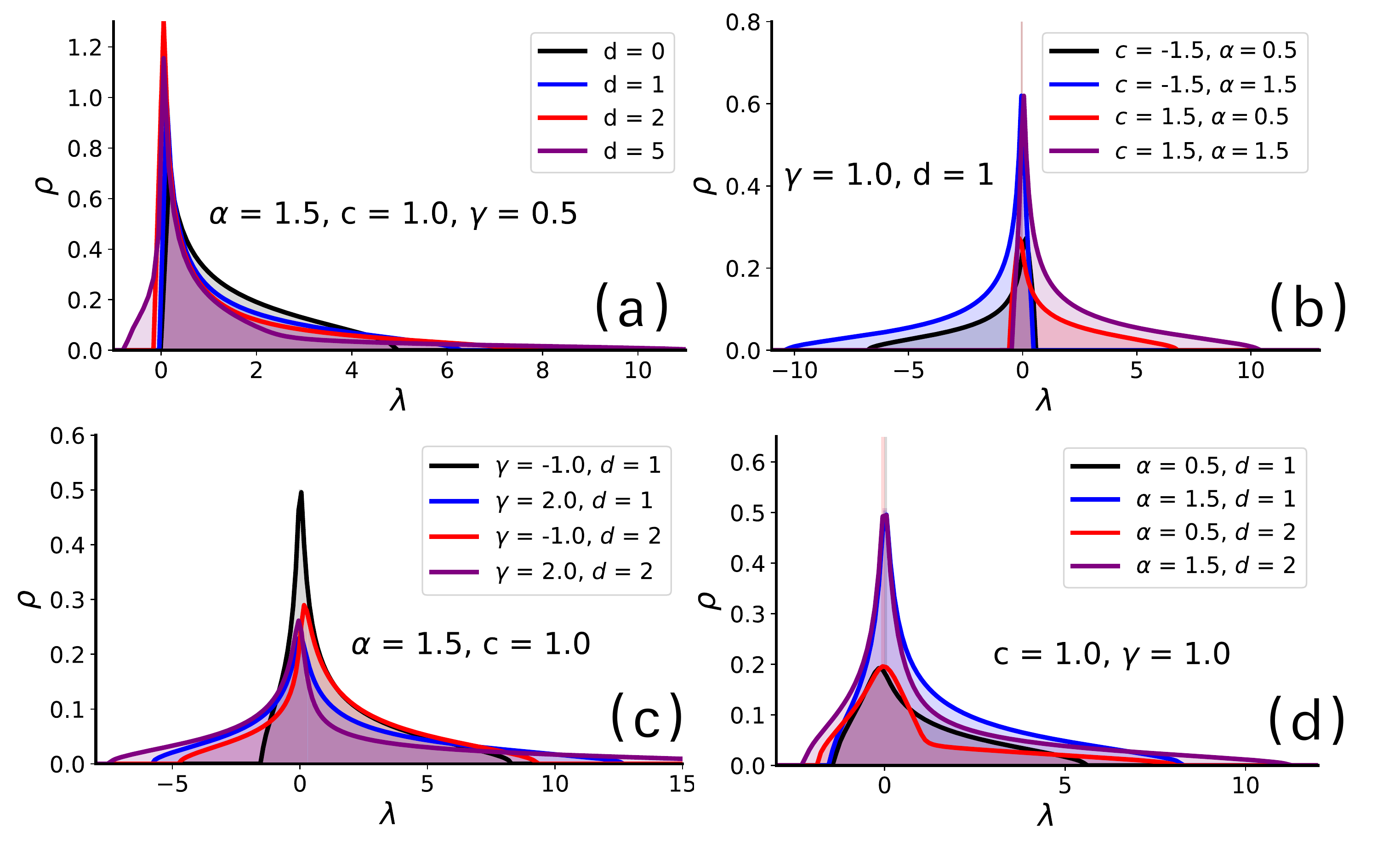}
  \caption{
  (Color online) Eigenvalue spectrum of $\J$ for different model parameters $c$, $d$, $\gamma$, and $\alpha$. 
	Solid lines are results from replica computation, and shadows are the numerical results
	of the eigenvalue spectrum of $1000\times 1000$ matrix averaged over $20$ instances. 
	(a) Spectrum of $\J$ for $\alpha$ = $1.5$, $c$ = $1.0$, $\gamma=0.5$ with different values of $d$.  
	(b) Spectrum of $\J$ for $d$ = $1$, $\gamma=1.0$ with different values of $\alpha$ and $c$.
	(c) Spectrum of $\J$ for $\alpha$ = $1.5$ and $c=1$ with different values of $\gamma$ and $d$. 
	(d) Spectrum of $\J$ for $c=1.0$ and $\gamma=1.0$ with different values of $\alpha$ and $d$.
  }\label{density}
\end{figure}
\section{Results}
In general, solving Eq.~(\ref{eq:spetrum3}) in an analytic form is very challenging. When $d=0$ (or $\gamma=0$), the spectrum equation can be
analytically solved, yielding the well-known result of Marchenko-Pastur law~\cite{MPlaw,Huang-2020weak}. We next give some simple arguments about the 
case of $d=1$.

When $d=1$, we have $A(x)= c+2\gamma \cos(2\pi x)$.
 First we rewrite Eq.~(\ref{eq:spetrum3}) as
\begin{equation}
\begin{aligned}
(1-\alpha)\frac{1}{G_{\mathbf{J}}}-\lambda &= -\frac{\alpha}{G_{\mathbf{J}}}\frac{1}{2\pi}\int _0^{2\pi} \mathrm{d}x\frac{1}{1-G_{\mathbf{J}}A(\frac{x}{2\pi})}\\
& = -\frac{\alpha}{G_{\mathbf{J}}}\frac{1}{2\pi}\int_0^{2\pi} dx \frac{1}{1-G_{\mathbf{J}}c-2G_{\mathbf{J}}\gamma\cos(x)}.
\label{eq:spetrum4}
\end{aligned}
\end{equation}
By substituting $\cos(x)=\frac{\mathrm{e}^{\i x}+\mathrm{e}^{-\i x}}{2}$ and
introducing a new integral variable $z=e^{\i x}$, we have
\begin{equation}
(1-\alpha)\frac{1}{G_{\mathbf{J}}}-\lambda = \f{\alpha}{G_\J}\frac{1}{\pi \i}\frac{1}{2 G_{\mathbf{J}}\gamma}\int_{|z|=1}dz\frac{1}{(z-z_0)(z-z_1)},
\end{equation}
where $z_0, z_1$ are the roots of $z^2 -2\frac{1-G_{\mathbf{J}} c}{2G_{\mathbf{J}} \gamma}z+1=0$, which implies
that $z_0 z_1 =1$ and $z_0+z_1=\frac{1-G_{\mathbf{J}} c}{G_{\mathbf{J}} \gamma}$. A simple argument leads to the conclusion that
only one of $(z_0,z_1)$ falls in the contour $|z|=1$.
Assuming $z_0$ (or $z_1$) is inside the contour, we have:
\begin{equation}
(1-\alpha)-\lambda G_{\mathbf{J}} = \frac{\alpha}{G_{\mathbf{J}}\gamma}\frac{1}{z_0-z_1},
\label{eq:spectrum5}
\end{equation}
with 
\begin{equation}
z_0-z_1 = \pm2\sqrt{\left(\frac{1-G_{\mathbf{J}} c}{2\gamma G_{\mathbf{J}}}\right)^2-1}.
\end{equation}
It then follows that Eq.~(\ref{eq:spectrum5}) (squaring both sides) becomes
\begin{equation}
\left(\lambda^{2} G_{\mathbf{J}}^{2}-2(1-\alpha) \lambda G_{\mathbf{J}}+(1-\alpha)^{2}\right)\left(1+\left(c^{2}-4 \gamma^{2}\right) G_{\mathbf{J}}^{2}-2 c G_{\mathbf{J}}\right)-\alpha^{2}=0.
\end{equation}
A solution of $G_\J$ provides the eigen-spectrum. Generalization to $d>1$ is straightforward, but an analytic solution becomes much more complicated.

\begin{figure}
     \includegraphics[bb=13 13 704 566,width=0.8\textwidth]{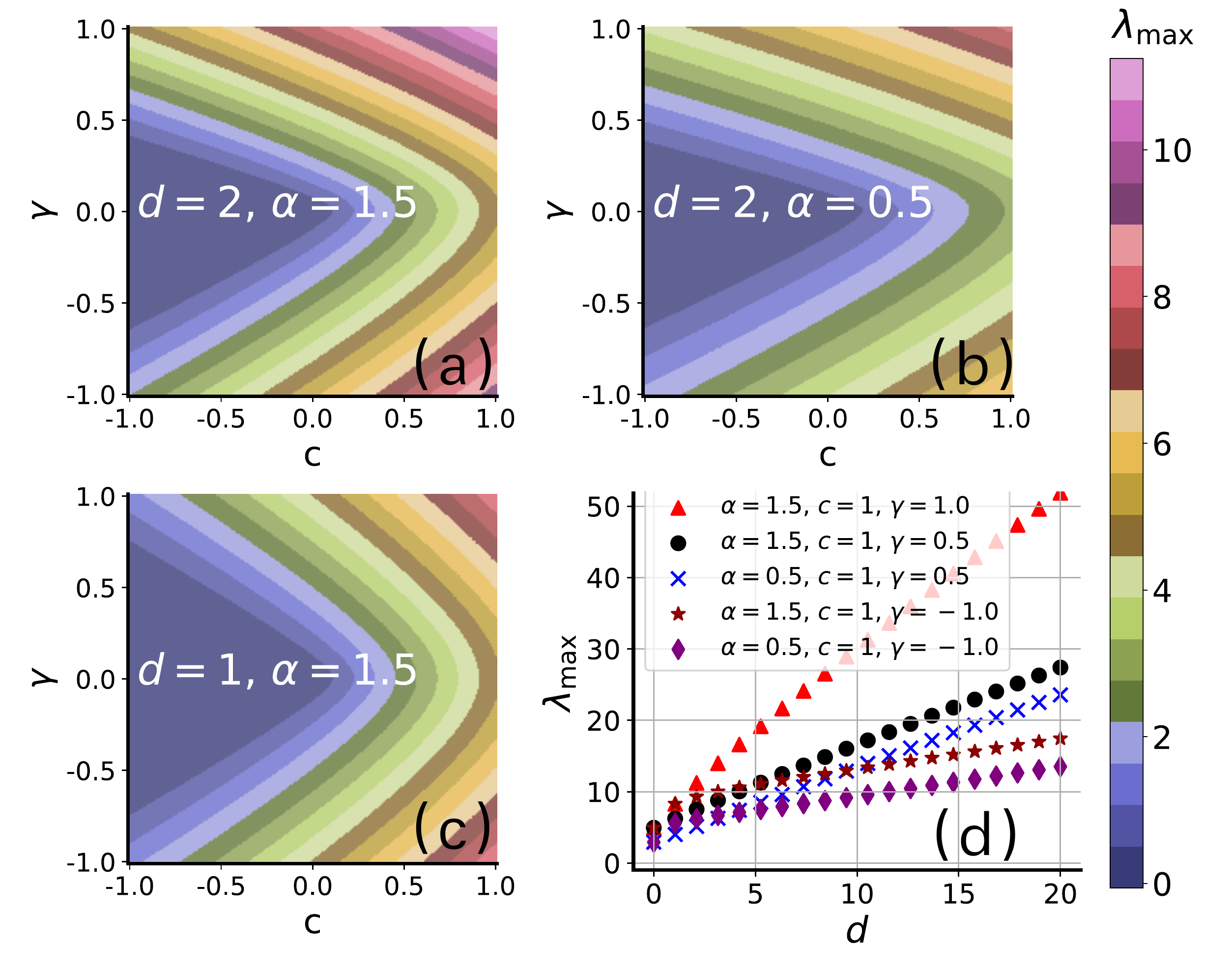}
  \caption{
  (Color online) The maximum eigenvalue of $\J$ for different model parameters $c$, $d$, $\gamma$, and $\alpha$. 
	(a) $d=2$ and $\alpha=1.5$. (b) $d=2$ and $\alpha=0.5$. (c) $d=1$ and $\alpha=1.5$. 
	(d) Relationship between the maximum eigenvalue and $d$ 
	under different values of $c$, $\gamma$, and $\alpha$.
  }\label{Maxeg}
\end{figure}

A numerical solution of the eigen-spectrum equation
can be obtained by separately solving the imaginary part and the real part of Eq.~(\ref{eq:spetrum3})
or doing a fixed-point iteration of Eq.~(\ref{eq:spetrum3}) in the complex domain. 
In practice, both strategies yield the identical result. 

By construction, $\rank(\J^\mu)=1$, and we then have $\rank(\J)\leq\sum_{\mu=1}^P\rank(\J^\mu)=P$. Alternatively, assuming that $X$ is a full-rank matrix, we have
\begin{equation}
	\rank(\J)=\rank\l(\frac{1}{N} \bxi^\T \bX \bxi\r) = \rank(\bxi)=\min{(P,N)} \ ,
\end{equation}
which means that, when $\a<1$, there are $N-P$ zero eigenvalues in the eigen-spectrum of $\J$.
Hence we write down the explicit spectrum considering the delta peak when $\a<1$ as follows
 \begin{equation}
 \rho(\lambda) = \begin{cases}
 \frac{1}{\pi}\im G_\J(\lambda) + (1-\alpha)\delta(\lambda),&\;\alpha<1;\\
 \frac{1}{\pi}\im G_\J(\lambda),&\;\alpha\ge1.
 \end{cases}
 \end{equation}
 Note that, $\frac{1}{\pi}\im G_\J(\lambda)$ gives the part of the spectrum density at $\lambda\not=0$.

\begin{figure}
     \includegraphics[bb=7 16 798 674,width=0.8\textwidth]{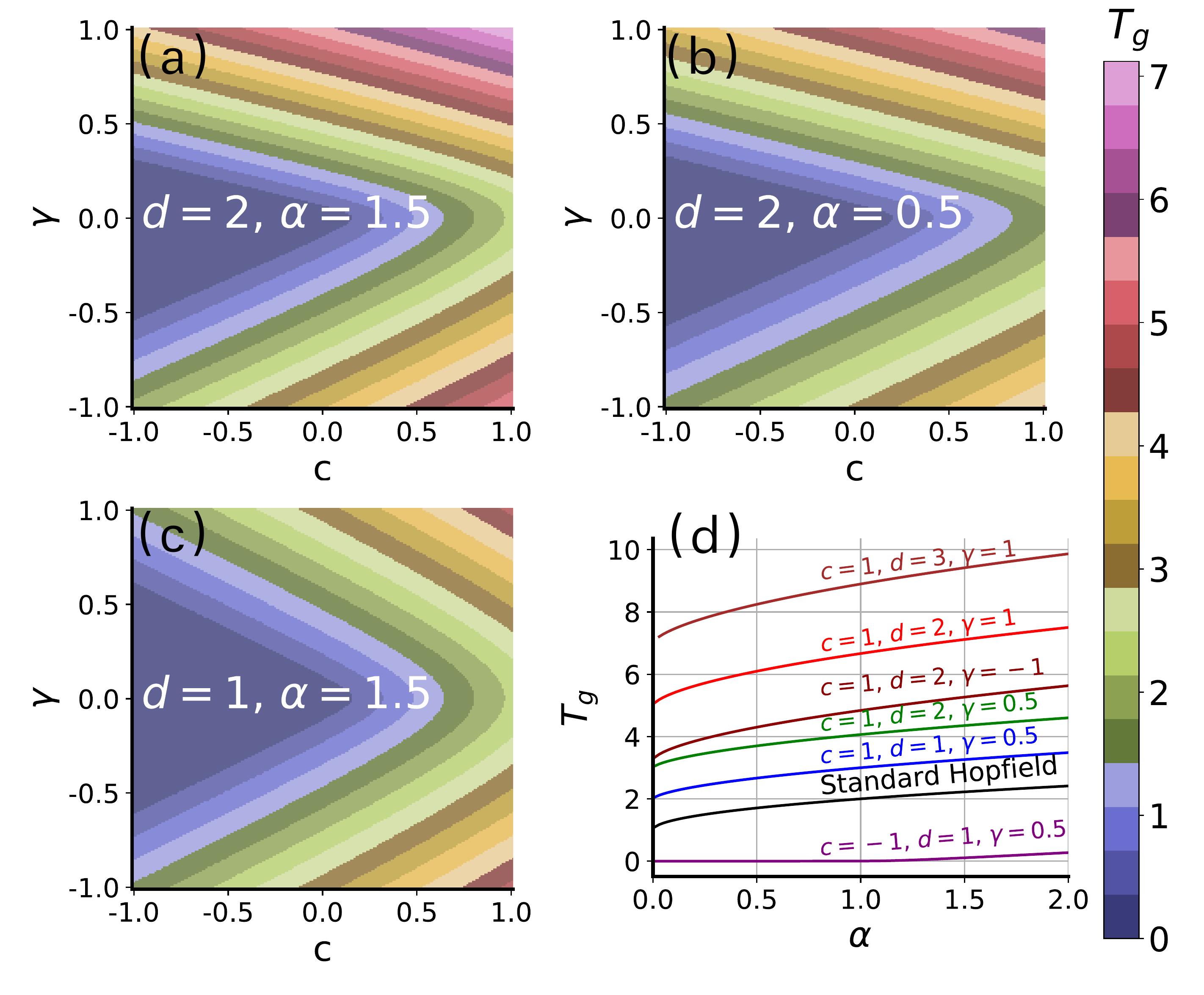}
  \caption{
  (Color online) The critical temperature $T_{g}$ where the paramagnetic phase to spin-glass phase transition occurs
	under different values of $c$, $d$, $\gamma$, and $\alpha$.  
	(a) $d=2$ and $\alpha=1.5$. (b) $d=2$ and $\alpha=0.5$. (c) $d=1$ and $\alpha=1.5$. 
	(d) Relationship between $T_g$ and $\alpha$ 
	under different values of $c$, $\gamma$, and $d$.
  }\label{transition}
\end{figure}

Comparison between the theory predictions and numerical simulations is plotted in Fig.~\ref{density}. We first consider the effects of 
the Hebbian length $d$. The case of $d=0$ corresponds to the standard Hopfield model, and the eigen-spectrum is the well-known Marchenko-Pastur law.
Increasing the value of $d$ strongly modifies the shape of the spectral density. More precisely, a 
large Hebbian length shapes a long tail, and there appear negative eigenvalues as well [Fig.~\ref{density} (a)]. 
Given the values of $\gamma$ and $d$, increasing the value of $\alpha$ stretches the tail of
the density profile. Moreover, the density profile is symmetric with respect to the origin
point ($\lambda=0$), for values of $c$ with the same magnitude but different signs. The extended 
correlation-span observed in the recent work~\cite{Fukai-2019} for negative values 
of $c$ (so called anti-Hebbian terms) may be related to the reconfiguration of the 
eigen-spectrum profile. A salient feature is that, negative eigenvalues become much 
more likely than the positive ones, when $c<0$ [see Fig.~\ref{density} (b) for $d=1$].

Increasing the Hebbian length also changes significantly the profile for 
negative values of $\gamma$, which allocates more density for negative eigenvalues [Fig.~\ref{density} (c)].
Negative $\gamma$ corresponds to introducing unlearning effects, which could remove
some of the original dominant attractors at positive $\gamma$ (no anti-Hebbian effects). 
Therefore, the unlearning term could reshape the energy landscape of the model~\cite{Huang-2021},
which may be connected to the specific shape of the eigen-spectrum. Given the same value
of $\alpha$ (memory load), the Hebbian length could also change the properties of the spectrum, e.g., 
large $d$ increases the eigenvalue span [Fig.~\ref{density} (d)].

Maximal eigenvalues are related to the stability of the paramagnetic phase, i.e., determining the transition temperature where
the paramagnetic phase is destablized toward the spin glass phase.
The impact of model parameters on the maximal eigenvalue is shown in Fig.~\ref{Maxeg}. 
For $d=1$, the value of $\lambda_{\rm max}$ is symmetric about the line at $\gamma=0$ [Fig.~\ref{Maxeg} (c)], 
while $d=2$ breaks this symmetry [Fig.~\ref{Maxeg} (a,b)].
This is due to the gauge invariance under the changes of $\gamma\to-\gamma$ and $\bx^\mu\to-\bx^\mu$ with odd (or even) indices,
which keeps the eigenvalue distribution invariant when $d=1$ but not when $d=2$.
In particular, large $\alpha$ 
increases the value of $\lambda_{\rm max}$, as expected from Fig.~\ref{density}. As shown
in Fig.~\ref{Maxeg} (d), the Hebbian length can affect the maximum eigenvalue. More precisely,
with increasing Hebbian length, the maximal eigenvalue grows in different manners, e.g.,
the strength of non-concurrent Hebbian terms is able to make the growth more rapid with the 
value of $d$. It is worth noticing that negative values of $\gamma$ (unlearning or anti-Hebbian
non-concurrent coupling) are able to significantly lower down the maximal eigenvalue.
Furthermore, Fig.~\ref{Maxeg} (d) shows that increasing $\alpha$ enhances $\lambda_{\rm max}$ in both cases of 
positive and negative $\gamma$.
Altogether, 
the Hebbian length in our associative memory model can have a significant impact on the spectral
density and hence the maximal eigenvalue.

\begin{figure}
    \centering
    \includegraphics[bb=10 12 423 283,width=0.6\textwidth]{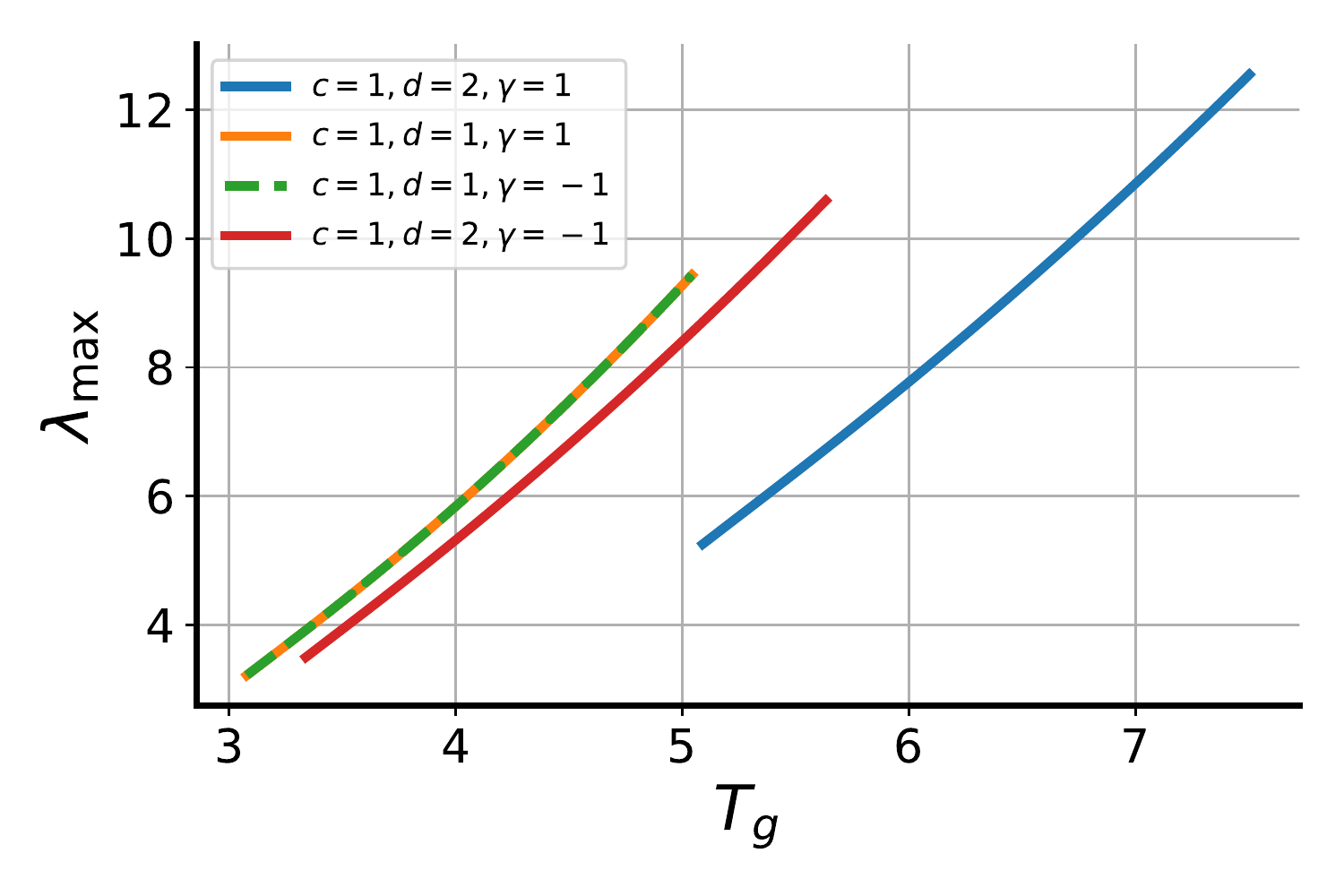}
    \caption{Relationship between the critical temperature $T_g (\alpha)$ 
    and the maximum eigenvalue $\lambda_{\max} (\alpha)$. 
    The memory load $\alpha$ changes from $0$
    to $2$. 
    Each point of the curve is obtained by solving corresponding iterative equations
    under specific parameter settings of $(\alpha,c,d,\gamma)$. }
    \label{templam}
\end{figure}

We finally study the transition from the paramagnetic phase to the
spin glass phase, whose precise location can be determined by our theory
[Eq.~(\ref{eq:beta_g})]. As shown in Fig.~\ref{transition}, $d=2$ breaks the symmetry about 
the origin ($\gamma=0$), as also expected from the profile of $\lambda_{\rm max}$. As $d$ 
increases, $T_{g}$ increases as well. As a consequence, the paramagnetic phase shrinks with 
increasing Hebbian length. By changing the sign of the non-concurrent Hebbian strength from 
being positive to being negative, the transition temperature is significantly lowered down.
By changing the sign of the concurrent Hebbian term in the same way, the paramagnetic phase 
could be significantly expanded as well. In this case, when $d=1$, non-concurrent Hebbian terms 
compete with the concurrent anti-Hebbian terms, leading to an expanded paramagnetic phase.
Below the critical line, there appear different kinds of phases, e.g., spin glass phase,
retrieval phase, correlated-attractor phase, and unlearning-modified correlated-attractor
phase, which are studied in detail in an accompany paper~\cite{Huang-2021}.
The relationship between the maximal eigenvalue and the transition temperature is shown in Fig.~\ref{templam}.

\section{Concluding remarks}
In conclusion, we derive the asymptotic spectral density of the associative
memory model with arbitrary Hebbian length, using different theoretical tools. 
In addition, the maximum eigenvalue equation can also be obtained by transforming
the original problem to a physics problem of computing the ground state, whose results coincide
with the vanishing-imaginary part of the Green's function. The maximum eigenvalue 
is then connected to the paramagnetic-to-glass transition. We finally remark that
the Hebbian length yields strong impacts on the above statistical properties of 
the model. We therefore transform an associative memory problem to a mathematical
problem going beyond the classic  Marchenko-Pastur law. Our study  would hopefully
provide further insights for the biological learning in wide integration windows, 
and even unlearning effects in reshaping the memory landscape during dreaming.


\begin{acknowledgments}
This research was supported by the National Natural Science Foundation of China for
Grant No. 11805284 (HH) and the start-up budget 74130-18831109 of the 100-talent-program 
of Sun Yat-sen University (HH), and research grants council of Hong Kong (grant numbers 16302419 and 16302619) (MW). 
\end{acknowledgments}



\end{document}